\documentclass[
  prd,preprint,longbibliography,
  showpacs,showkeys,lengthcheck,
  nofootinbib,tightenlines,onecolumn,notitlepage,
  preprintnumbers,superscriptaddress
]{revtex4-1}

\usepackage{graphicx}
\usepackage[usenames,dvipsnames]{xcolor}
\graphicspath{{figures/}}
\usepackage{tikz-feynman}

\usepackage{hyperref}
\hypersetup{colorlinks=true,citecolor=blue,linkcolor=blue,urlcolor=blue}

\usepackage{diagbox}
\usepackage{booktabs}

\usepackage{amsmath,amssymb,amsfonts}
\usepackage{mathrsfs}
\usepackage{slashed}
\usepackage{bm}

\usepackage[final]{changes} 

\usepackage[utf8]{inputenc}
\usepackage{newtxtext}
\usepackage[mathcal]{euscript}

\newcommand{\Lag}{\ensuremath{\mathscr{L}}}
\newcommand{\dx}{\ensuremath{\xi}}
\newcommand{\Lie}{\ensuremath{\mathcal{L}_{\dx}}}
\newcommand{\Tmunu}{\ensuremath{T^{\mu}_{\phantom{\mu}\nu}}}
\newcommand{\Dmunu}{\ensuremath{\mathscr{D}^{\mu}_{\phantom{\mu}\nu}}}
\newcommand{\gmunu}{\ensuremath{\delta^{\mu}_{\phantom{\mu}\nu}}}

\begin{document}

\title{Noether's theorems and the energy-momentum tensor in quantum gauge theories}

\author{Adam Freese}
\email{afreese@uw.edu}
\affiliation{Department of Physics, University of Washington, Seattle, WA 98195, USA}

\begin{abstract}
  Noether's first and second theorems both imply conserved currents
  that can be identified as an energy-momentum tensor (EMT).
  The first theorem identifies the EMT as the conserved current associated
  with global spacetime translations,
  while the second theorem identifies it as a conserved current
  associated with local spacetime translations.
  This work obtains an EMT for quantum electrodynamics and quantum chromodynamics
  through the second theorem,
  which is automatically symmetric in its indices and invariant under the
  expected symmetries (e.g., BRST invariance)
  without the need for introducing an ad hoc improvement procedure.
\end{abstract}

\preprint{NT@UW-21-20}

\maketitle


\section*{Erratum}

This paper was erroneous in its treatment of spinor fields.
In particular, it assumed that spinor fields transformed under a double cover
of the group $\mathrm{GL}(4,\mathbb{R})$ under general coordinate transformations,
which when restricted to the subgroup of Lorentz transformations
would reproduce the usual Lorentz transformation formula for spinor fields.
The formulas in Eq.~(\ref{eqn:variations:spinor}) were obtained in this manner.

However, a finite linear double cover of $\mathrm{GL}(4,\mathbb{R})$ does not exist.
This has long been known, with the classic proofs given by
Weyl~\cite{Weyl:1929fm} and Cartan~\cite{Cartan:1966}.
For this reason, spinor fields are typically considered to transform like
scalar fields under general coordinate transformations.
The spinorial properties of spinor fields are instead accounted for within
the tetrad or vierbein formalism,
which is explained in-depth in
Appendix J of Carroll~\cite{Carroll:2004st},
Chapter 11 of Hamilton~\cite{hamilton2018general},
and Chapter 12 of Weinberg~\cite{Weinberg:1972kfs}.

In light of this, Eq.~(\ref{eqn:variations:spinor})
does not correctly describe the transformation of spinor fields
under local translations.
This leads to a fork in the road:
either Eq.~(\ref{eqn:variations:spinor}) must be modified
to remove the $(\partial_\alpha \xi_\beta) \sigma^{\alpha\beta}$ terms,
or the local translation must be supplemented with an additional
internal transformation of the vierbein.

The former case
(having spinor fields transform like scalar fields)
is explored in Ref.~\cite{Freese:2025glz},
which additionally gives a clearer exposition of the procedure
of this work and an elementary proof that there is
no finite linear double cover of $\mathrm{GL}(4,\mathbb{R})$.
The end result is an energy-momentum tensor for quantum chromodynamics
that is gauge-invariant but asymmetric---%
and which coincides with the gauge-invariant kinetic energy-momentum tensor
of Leader and Lorc\'e~\cite{Leader:2013jra}.

The latter case
(supplementing local translations with an internal vierbein transformation)
is explored by another recent work~\cite{BarroseSa:2025uxe}.
This combined transformation does produce the symmetric energy-momentum tensor,
but this procedure differs from a pure spacetime translation.

The remainder of this document is unmodified,
including erroneous steps and conclusions.
Please see Ref.~\cite{Freese:2025glz}
for an alternate derivation in which the errors in this work are absent.


\section{Introduction}
\label{sec:intro}

The energy-momentum tensor (EMT) has recently become a major focus of both
theoretical and experimental efforts in hadron physics.
It is widely believed that a deeper understanding of the EMT
in quantum chromodynamics (QCD)
and empirical knowledge of its matrix elements for physical hadron states
will illuminate major puzzles such as the origin and decomposition of the proton's
mass~\cite{Ji:1994av,Ji:1995sv,Lorce:2017xzd,Hatta:2018sqd,Metz:2020vxd,Ji:2021mtz,Lorce:2021xku}
and the distribution of its
spin~\cite{Ashman:1987hv,Jaffe:1989jz,Ji:1996ek,Leader:2013jra}.

Despite this concerted focus,
the EMT of QCD appears to be constructed on unsure ground.
The so-called canonical EMT is defined as
the conserved current associated with global spacetime translation symmetry
via Noether's first theorem~\cite{Noether:1918zz}.
However, the canonical EMT lacks symmetry properties of the underlying theory,
such as conformal invariance and gauge or BRST invariance.
This appears to be a defect with Noether's first theorem,
and the situation is typically rectified through an ad hoc
Belinfante improvement procedure~\cite{Belinfante:1939emt}.

The Belinfante improvement procedure produces an EMT
with the desired symmetry properties
by adding the divergence of a superpotential to the canonical result.
This additional term is defined so that its divergence trivially vanishes,
meaning it can be added to the EMT multiplied by any factor
and produce a conserved current
(though one that will be asymmetric in general).
The freedom of choice for the superpotential
gives one the liberty to obtain different EMT operators
which produce different results for hadronic matrix elements.
This is not merely a hypothetical concern:
different versions of the QCD EMT operator exist in the literature
(see Ref.~\cite{Leader:2013jra} for a review).
Matrix elements of different EMT operators can differ in the number
of form factors they produce in the matrix elements of hadronic
states~\cite{Lorce:2018egm,Cosyn:2019aio}
or even in what values these form factors take~\cite{Hudson:2017xug}.
Additionally, the canonical and Belinfante-improved EMTs
entail different angular momentum densities for electromagnetic
fields~\cite{Jaffe:1989jz,Ji:1996ek},
which may open the question of which EMT is physically correct
to experimental determination with an optical measurement~\cite{Leader:2017htb}.
This is clearly not an ideal situation;
it would be greatly preferable to have an unambiguous method for
arriving at a predetermined EMT operator from the start.

Such an unambiguous method does in fact exist:
it is the direct application of Noether's second theorem,
which pertains to local symmetries,
such as local gauge transformations and local translations.
In fact, several popular quantum field theory
textbooks~\cite{Itzykson:1980rh,Weinberg:1995mt}
approach the derivation of canonical the energy-momentum tensor
through local coordinate transformations rather than global transformations,
but apply the approach inconsistently by
treating the components of non-scalar fields as
a collection of scalar fields under these local translations.
Several works over the last few decades have found that
accounting for the correct transformation properties of vector fields
when considering local spacetime translations
gives an EMT operator with the expected symmetry properties for classical
electromagnetism~\cite{Felsager:1981iy,Munoz:1996wp,GamboaSaravi:2002vos,Montesinos:2006th},
classical Yang-Mills theory~\cite{GamboaSaravi:2003aq,Montesinos:2006th},
and classical Proca theory~\cite{Montesinos:2006th}.
In fact, an early and neglected work of Bessel-Hagen~\cite{Bessel:1921emt}
found the correct physical EMT for electromagnetism as early as 1921.

The findings of
Refs.~\cite{Bessel:1921emt,Felsager:1981iy,Munoz:1996wp,GamboaSaravi:2002vos,GamboaSaravi:2003aq,Montesinos:2006th}
unfortunately remain obscure in the hadron physics community,
and one of my purposes with this manuscript is to rectify this.
This paper additionally contains new findings:
I use Noether's second theorem to directly derive the correct
EMT operator---with the expected symmetry properties---for theories
that include spinor fields,
including quantum electrodynamics (QED) and QCD.

This work is organized as follows.
Sec.~\ref{sec:formalism} opens with a derivation of a general formula
for the EMT obtained through Noether's second theorem.
Sec.~\ref{sec:illustrate}
illustrates the application of this general formula
to several simple free field theories;
this is primarily a didactic section,
with the scalar and vector results existing elsewhere in the literature.
I obtain the EMT operators for both QED and QCD
in Sec.~\ref{sec:gauge},
and Sec.~\ref{sec:conclusion} then concludes the work.


\section{Derivation of the energy-momentum tensor}
\label{sec:formalism}

In Ref.~\cite{Noether:1918zz},
Noether proved two theorems related to symmetries of the action.
Her first theorem entails the existence of a conserved current associated
with any continuous symmetry of the action,
while her second theorem applies only to infinite-dimensional groups.
In particular, Noether's second theorem can be used to derive conserved currents
if the symmetry of the action is encoded by an arbitrary function
(or set of functions), rather than a finite set of parameters.
The crucial difference between the theorems is that surface integrals
are not dropped in the first theorem, but they are in the second theorem.
For careful treatments and a full exposition of the two theorems,
see Refs.~\cite{Brading:2000hc,Kosyakov:2007qc,DeHaro:2021gdv},
as well as Noether's original paper~\cite{Noether:1918zz}.

The canonical energy-momentum tensor (EMT) is obtained by applying
Noether's \emph{first} theorem to global spacetime translations,
i.e. $x \mapsto x + \dx$ for constant $\dx$.
The well-known result is~\cite{Itzykson:1980rh,Peskin:1995ev,Weinberg:1995mt,Kosyakov:2007qc}:
\begin{align}
  \label{eqn:emt:canonical}
  T^{\mu\nu}(x)
  =
  \sum_i
  \frac{\partial\Lag}{\partial (\partial_\mu \phi_i)} \partial^\nu \phi_i
  -
  g^{\mu\nu} \Lag
  \,,
\end{align}
where the sum is over all fields in the theory,
and for non-scalar fields (e.g.\ vector fields)
field components are also summed over.
The canonical EMT for non-scalar fields is asymmetric in its indices,
and for gauge theories is infamously gauge dependent.
Both issues are conventionally fixed using the Belinfante
improvement procedure~\cite{Belinfante:1939emt}.

The standard quantum field theory textbooks by
Itzykson and Zuber~\cite{Itzykson:1980rh}
and Weinberg~\cite{Weinberg:1995mt}
covertly employ Noether's second theorem rather than her first
to derive the energy-momentum tensor,
since they use spacetime-dependent translations
$x \mapsto x + \dx(x)$
and drop surface integrals in the course of the derivation.
These derivations lead to the same EMT for scalar fields as the canonical
derivation via Noether's first theorem, namely Eq.~(\ref{eqn:emt:canonical}).
The authors then assume this result likewise applies to vector and spinor fields,
but this assumption is incorrect, as I will show below.
In fact, by using Noether's second theorem and carefully accounting
for the transformation properties of the fields,
the resulting EMT will automatically be symmetric in its indices,
and for gauge theories will also be gauge invariant.

\begin{figure}
  \begin{align*}
    \vcenter{\hbox{
      \includegraphics[width=0.24\textwidth]{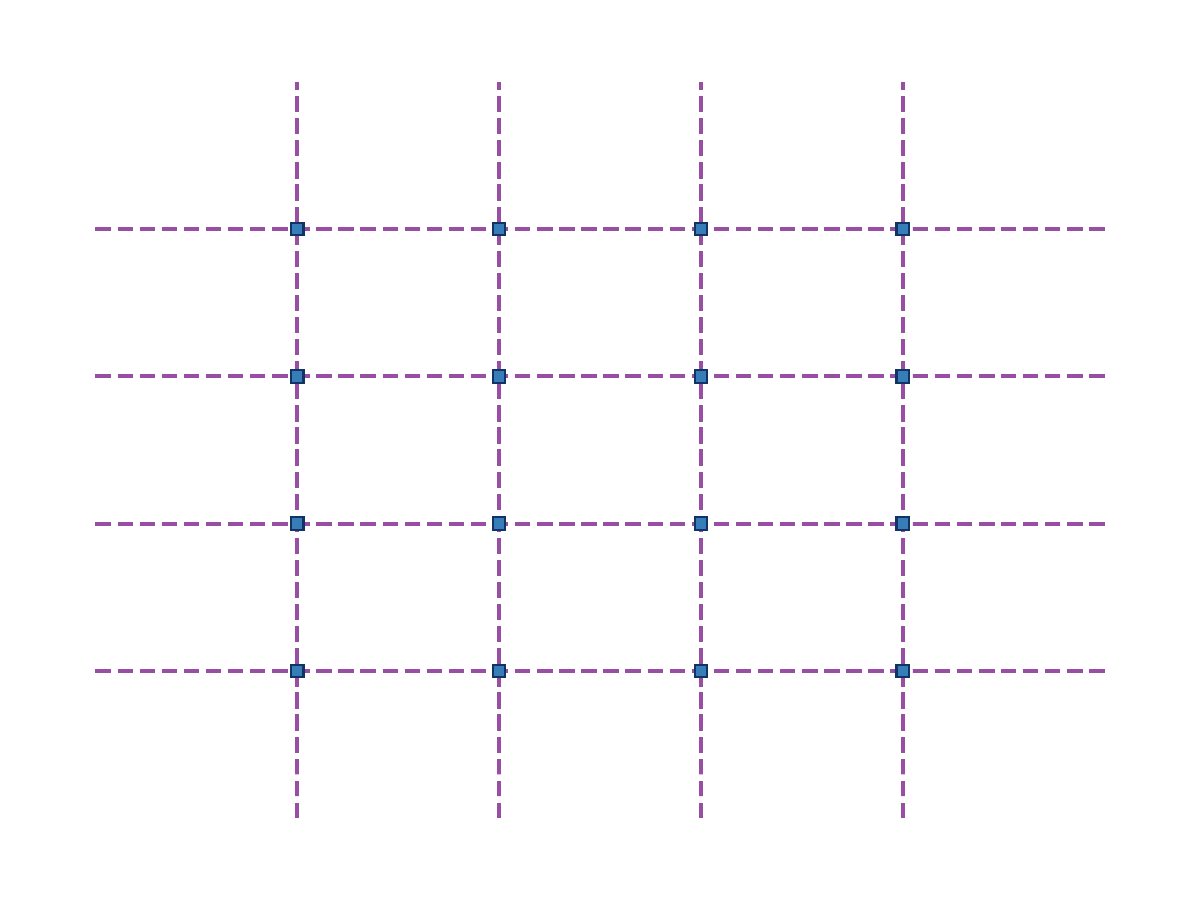}
    }}
    \xrightarrow[x \mapsto x + \dx(x)]{}
    \vcenter{\hbox{
      \includegraphics[width=0.24\textwidth]{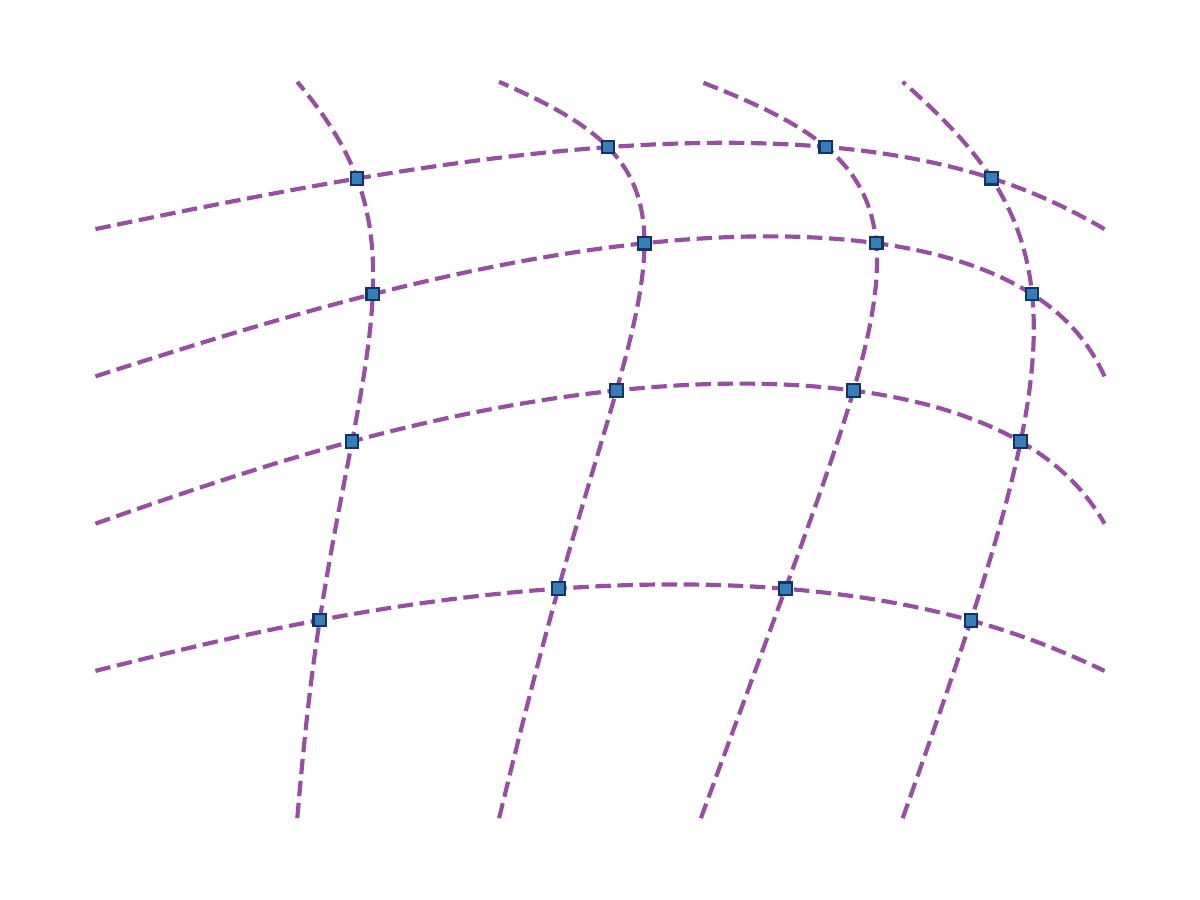}
    }}
  \end{align*}
  \caption{
    Depiction of a local spacetime translation.
  }
  \label{fig:translation}
\end{figure}

Let us proceed to show how an energy-momentum tensor
can be obtained using Noether's second theorem.
The transformation employed is an infinitesimal spacetime-dependent translation:
\begin{subequations}
  \label{eqn:translation}
  \begin{align}
    x^\mu
    \mapsto
    x'^\mu
    =
    x^\mu + \dx^\mu(x)
    \,,
  \end{align}
  which is depicted visually in Fig.~\ref{fig:translation}.
  In addition to spacetime being translated,
  the total variation of any scalar field must remain unchanged:
  \begin{align}
    \phi'(x')
    =
    \phi(x)
    \,.
  \end{align}
  The last stipulation is that the total variation of the metric tensor is zero:
  \begin{align}
    \label{eqn:metric}
    g'_{\mu\nu}(x')
    =
    g_{\mu\nu}(x)
    \,.
  \end{align}
  This last stipulation is made to avoid any discussion
  of how the action depends on the metric, and therefore of gravitation.
  It additionally makes the local translation a physical transformation
  rather than a mathematically trivial reparametrization.
\end{subequations}
Recall that in Minkowski spacetime,
the metric $g_{\mu\nu} = g^{\mu\nu} = \mathrm{diag}(1,-1,-1,-1)$
is invariant under Lorentz transformations and global translations.

The condition on scalar fields requires that the functional form of the scalar field changes:
\begin{align}
  \phi'(x')
  =
  \phi'(x + \dx(x))
  \approx
  \phi'(x)
  +
  \dx^\lambda(x) \partial_\lambda \phi(x)
  +
  \mathcal{O}(\dx^2)
  \,.
\end{align}
Thus the total variation of the field is zero, but its functional variation is not.
\begin{subequations}
  \begin{align}
    \Delta \phi
    & \equiv
    \phi'(x')
    -
    \phi(x)
    =
    0
    \\
    \delta \phi
    & \equiv
    \phi'(x)
    -
    \phi(x)
    =
    - \dx^\lambda(x) \partial_\lambda \phi(x)
    \,.
  \end{align}
\end{subequations}
This assumption has implications for how non-scalar fields transform.
For the derivative of a scalar field, for instance:
\begin{align}
  \Delta (\partial_\mu \phi)
  \equiv
  \frac{\partial \phi'(x')}{\partial x'^\mu}
  -
  \frac{\partial \phi(x)}{\partial x^\mu}
  =
  \frac{\partial \phi(x)}{\partial x'^\mu}
  -
  \frac{\partial \phi(x)}{\partial x^\mu}
  =
  -(\partial_\mu \dx^\lambda(x)) \partial_\lambda \phi(x)
  +
  \mathcal{O}(\dx^2)
  \,.
\end{align}
Just as with the familiar case of Lorentz transformations,
$\partial_\mu\phi$
is said to transform covariantly---and likewise is any other field
that obeys the transformation rule $\Delta A_\mu = -(\partial_\mu\dx^\lambda(x))A_\lambda(x)$.
Under Lorentz transforms, a field with an upper index $B^\mu$
transforms contravariantly,
such that $A_\mu B^\mu$ is a scalar and remains invariant.
As usual, the metric can be used to raise and lower indices,
such as in $A^\mu = g^{\mu\nu}A_\nu$.
\replaced{
However, under the local translations considered in this work,
}{
Since
the transformation of Eq.~(\ref{eqn:translation})
does not alter the metric,
}
a field that transforms covariantly under Eq.~(\ref{eqn:translation})
\replaced{
does not in general transform contravariantly when its index is raised.
}{
still transforms covariantly even if its index is raised.
}
For instance:
\begin{align*}
  \Delta (\partial^\mu \phi)
  =
  \Delta (g^{\mu\nu} \partial_\nu \phi)
  =
  g^{\mu\nu} \Delta( \partial_\mu \phi)
  =
  - (\partial^\mu \dx^\lambda(x)) \partial_\lambda \phi(x)
  +
  \mathcal{O}(\dx^2)
  \,.
\end{align*}
Another consequence of this is that the contraction of a field with itself
is not \added{in general} invariant under Eq.~(\ref{eqn:translation}).
For instance, if $A_\mu(x)$ is a covariant field then:
\begin{align*}
  \Delta (A^\mu A_\mu)
  =
  A^\mu(x)
  (\Delta A_\mu)
  +
  (\Delta A^\mu) A_\mu(x)
  +
  \mathcal{O}(\dx^2)
  =
  - 2 g^{\mu\nu}
  A_\mu(x)
  (\partial_\nu \dx^\lambda(x))
  A_\lambda(x)
  +
  \mathcal{O}(\dx^2)
  \,.
\end{align*}
This is in stark contrast to the familiar case of Lorentz transforms,
for which $A_\mu A^\mu = g^{\mu\nu} A_\mu A_\nu$ is invariant.
\added{
For lack of a better term, I will say that $A^\mu = g^{\mu\nu} A_\nu$
transforms upper-covariantly when $A_\mu$ transforms covariantly.
For Lorentz transforms,
``upper-covariantly'' can be considered a synonym of ``contravariantly,''
but this is not the case for general local translations.
}

A field $V^\mu$ is said to transform contravariantly
under Eq.~(\ref{eqn:translation}) if, for any covariant vector field $A_\mu$,
the four-product $A_\mu V^\mu$ is invariant under local translations,
i.e., if $\Delta( V^\mu A_\mu) = 0$.
Since infinitesimal differences obey a product rule:
\begin{align*}
  0
  &=
  \Delta(V^\mu A_\mu)
  =
  V^\mu(x) (\Delta A_\mu)
  +
  (\Delta V^\mu) A_\mu(x)
  =
  -V^\mu (\partial_\mu \dx^\lambda(x)) A_\lambda(x)
  +
  (\Delta V^\mu) A_\mu(x)
  \\
  &=
  \Big(
  -V^\lambda(x) (\partial_\lambda \dx^\mu(x))
  +
  (\Delta V^\mu)
  \Big)
  A_\mu(x)
  \,,
\end{align*}
and since this is true for any covariant vector field $A_\mu(x)$ it follows that:
\begin{align}
  \Delta V^\mu(x)
  =
  (\partial_\lambda \dx^\mu(x)) V^\lambda(x)
  \,.
\end{align}
Again, since the metric remains unaltered by the transformation of
Eq.~(\ref{eqn:translation}), $V_\mu = g_{\mu\nu} V^\nu$
\replaced{
does not transform covariantly,
and for lack of a better term I say it transforms lower-contravariantly.
Note that under Lorentz transforms, ``lower-contravariantly''
can be considered a synonym of ``covariantly.''
}{
transforms contravariantly rather than covariantly.
}

In general, it is helpful (as a space-saving and bookkeeping measure)
to raise and lower indices at will, as is common practice in the
broader physics community.
As we have seen, this does not
\replaced{
turn covariant vector fields into contravariant vector fields
and vice-versa.
It is thus necessary to keep track of whether a vector field is a covariant field
(meaning it transforms covariantly when its index is lowered)
or a contravariant vector field
(meaning it transforms contravariantly when its index is raised).
}{
alter the transformation properties
of the field under Eq.~(\ref{eqn:translation}).
It is necessary to keep track of whether a vector field is
covariant or contravariant under local translations,
but this is not in general determined by whether its index is upper or lower.
}

The following key results summarize the transformation properties
obtained so far:
\begin{subequations}
  \label{eqn:variations}
  \begin{align}
    &
    \Delta \phi
    =
    0
    &
    \text{Scalar field}
    \phantom{\,.}
    \\
    &
    \Delta A_\mu
    =
    - (\partial_\mu \dx^\lambda(x))
    A_\lambda(x)
    &
    \text{Covariant vector field}
    \phantom{\,.}
    \\
    &
    \Delta V^\mu
    =
    (\partial_\lambda \dx^\mu(x))
    V^\lambda(x)
    &
    \text{Contravariant vector field}
    \phantom{\,.}
    \\
    &
    \Delta F_{\mu\nu}
    =
    -
    (\partial_\mu \dx^\lambda(x)) F_{\lambda\nu}
    -
    (\partial_\nu \dx^\lambda(x)) F_{\mu\lambda}
    &
    \text{Rank-2 covariant tensor field}
    \,.
  \end{align}
\end{subequations}
\added{
Rules for upper-covariant and lower-contravariant transformations
can be obtained by simply raising or lowering the indices in the rules given above.
}
The formula for the rank-2 covariant tensor field is helpful in particular
for derivatives of covariant vector fields, $\partial_\mu A_\nu(x)$.
It's worth noting that all of these variations are linear in first derivatives
of the translation function $\dx^\nu(x)$.

\added{
These transformation rules are peculiar,
but they are essentially a formalization of the local translation used by
Itzykson and Zuber~\cite{Itzykson:1980rh}
to obtain the EMT of a scalar field.
To help ground the reader a bit,
it is interesting to note that infinitesimal Lorentz transforms are a subset
of local translations when the translation field is given by:
\begin{align*}
  \dx^\mu(x)
  =
  \omega^{\mu}_{\phantom{\mu}\nu} x^\nu
  \,,
\end{align*}
where $\omega^{\mu}_{\phantom{\mu}\nu} = -\omega_{\nu}^{\phantom{\nu}\mu}$
is a small constant antisymmetric tensor.
Using Eq.~(\ref{eqn:variations}) with these infinitesimal Lorentz transforms,
covariant and upper-covariant transformations respectively can be written:
\begin{align*}
  \Delta A_\mu
  &=
  -\omega^{\nu}_{\phantom{\nu}\mu} A_\nu
  =
  \omega_{\mu}^{\phantom{\mu}\nu} A_\nu
  \\
  \Delta A^\mu
  &=
  -\omega_{\nu}^{\phantom{\nu}\mu} A^\nu
  =
  \omega^{\mu}_{\phantom{\mu}\nu} A^\nu
  \,,
\end{align*}
where the antisymmetry of $\omega^{\mu}_{\phantom{\mu}\nu}$ was used.
Similarly, contravariant and lower-contravariant transforms can be written:
\begin{align*}
  \Delta V^\mu
  &=
  \omega^\mu_{\phantom{\mu}\nu} V^\nu
  \\
  \Delta V_\mu
  &=
  \omega_\mu^{\phantom{\mu}\nu} V_\nu
  \,.
\end{align*}
Comparing these transformation rules,
it should be clear that in the case of infinitesimal Lorentz transforms,
the upper-covariant and contravariant transformation rules are the same,
and likewise the lower-contravariant and covariant rules are identical.
The antisymmetry of $\partial_\nu \dx^\mu(x) = \omega^\mu_{\phantom{\mu}\nu}$
is what allowed this to happen.
For general local translations, however, $\partial_\nu \dx^\mu(x)$ is not antisymmetric,
and for this reason it is necessary to distinguish between covariant and contravariant vector fields.
}

Another curiosity is that all of these variations can be expressed in terms of
a mathematical concept called the Lie derivative.
In particular, if $\Lie$ is the Lie derivative along $\dx^\mu(x)$,
then for any kind of field $\Psi$:
\begin{align}
  \label{eqn:delta:lie}
  \Delta \Psi
  =
  \dx^\lambda(x) \partial_\lambda \Psi(x)
  -
  \Lie[\Psi(x)]
\end{align}
under the transformation of Eq.~(\ref{eqn:translation}).
Formal definitions and expositions of the properties of Lie derivatives can be found in
Refs.~\cite{slebodzinski2010republication,trautman2010editorial,Dewitt:1978anl,Carroll:1997ar,Nakahara:2003nw,Yano:2020lie,Friedman:2013xza}.
I will not require the Lie derivative in the derivation to follow,
but since several works on the relation of the EMT with Noether's theorems
use them~\cite{GamboaSaravi:2002vos,GamboaSaravi:2003aq,Montesinos:2006th},
it is worth making note of how they can be related to this work.

Now I will proceed with the derivation.
Let $\Lag$ be a function of several fields $\Psi_l$.
The action is:
\begin{align}
  S
  =
  \int \mathrm{d}^4x \,
  \Lag[\Psi_l(x),\partial_\mu\Psi_l(x)]
  \,.
\end{align}
The transformation in Eq.~(\ref{eqn:translation}) is performed,
giving a new action:
\begin{align}
  S'
  =
  \int \mathrm{d}^4x' \,
  \Lag[\Psi'_l(x'),\partial'_\mu\Psi_l'(x')]
  \,.
\end{align}
The change in the action is just $\Delta S = S' - S$.
It is helpful to note that, to leading order in $\dx^\mu$,
the Jacobian of the variable change is:
\begin{align}
  \left| \frac{\partial (x'^0,x'^1,x'^2,x'^3) }{\partial (x^0,x^1,x^2,x^3) } \right|
  =
  1 + (\partial_\mu \dx^\mu)
  \,.
\end{align}
Thus the action variation is:
\begin{align}
  \Delta S
  =
  \int \mathrm{d}^4x \,
  \Big\{
    \Delta \Lag
    +
    (\partial_\mu \dx^\mu)
    \Lag
    \Big\}
  \,.
\end{align}
The chain rule is applied to the variation, giving:
\begin{align}
  \Delta S
  =
  \int \mathrm{d}^4x \,
  \left\{
    \sum_l \left(
    \frac{\partial\Lag}{\partial \Psi_l}
    \Delta \Psi_l
    +
    \frac{\partial\Lag}{\partial (\partial_\mu\Psi_l)}
    \Delta (\partial_\mu\Psi_l)
    \right)
    +
    (\partial_\mu \dx^\mu) \Lag
    \right\}
  \,.
\end{align}
Since $\Delta\Psi_l$ and $\Delta(\partial_\mu\Psi_l)$ are always linear in
derivatives of $\dx^\nu(x)$, I can define:
\begin{align}
  \label{eqn:D}
  \mathscr{D}^{\mu}_{\phantom{\mu}\nu}[\Psi_l]
  \equiv
  -
  \frac{\partial}{\partial(\partial_\mu\dx^\nu)}
  \left[
    \frac{\partial\Lag}{\partial \Psi_l}
    \Delta \Psi_l
    +
    \frac{\partial\Lag}{\partial (\partial_\mu\Psi_l)}
    \Delta (\partial_\mu\Psi_l)
    \right]
\end{align}
and write the action variation as:
\begin{align}
  \Delta S
  =
  \int \mathrm{d}^4x \,
  \left\{
    -
    \sum_l
    \mathscr{D}^{\mu}_{\phantom{\mu}\nu}[\Psi_l]
    (\partial_\mu \dx^\nu)
    +
    (\partial_\mu \dx^\mu) \Lag
    \right\}
  \,.
\end{align}
Now, I employ the symmetry assumptions behind Noether's second theorem.
Firstly, the transformation of Eq.~(\ref{eqn:translation}) is hypothesized to be a
symmetry of the action, so $\Delta S = 0$.
Secondly, the translation function $\dx^\nu(x)$ is assumed to be arbitrary,
aside from the requirement that it vanish outside some compact region of
spacetime~\cite{Noether:1918zz}.
This second assumption allows surface terms to be dropped when employing
integration by parts:
\begin{align}
  \int \mathrm{d}^4x \,
  \dx^\nu(x)
  \partial_\mu
  \left\{
    \sum_l
    \mathscr{D}^{\mu}_{\phantom{\mu}\nu}[\Psi_l]
    -
    \delta^{\mu}_{\phantom{\mu}\nu} \Lag
    \right\}
  =
  0
  \,,
\end{align}
and the arbitrariness of $\dx^\nu(x)$ means that the rest of the integrand must
always be zero:
\begin{align}
  \partial_\mu
  \left\{
    \sum_l
    \mathscr{D}^{\mu}_{\phantom{\mu}\nu}[\Psi_l]
    -
    \delta^{\mu}_{\phantom{\mu}\nu} \Lag
    \right\}
  =
  0
  \,.
\end{align}
Therefore,
\begin{align}
  \label{eqn:emt:general}
  \Tmunu(x)
  =
  \sum_l
  \Dmunu[\Psi_l]
  -
  \gmunu \Lag
\end{align}
is a conserved quantity,
and a candidate for the energy-momentum tensor.
Of course, since transformation properties under Eq.~(\ref{eqn:translation})
are unaltered by raising and lowering indices, we can raise the $\nu$
and use these equations to calculate $T^{\mu\nu}$ as the EMT.


\section{The energy-momentum tensor of simple theories}
\label{sec:illustrate}

In this section, I will consider several simple quantum field theories
and derive the second theorem EMT for each.
This is meant primarily as a didactic warm-up exercise
before I proceed to QED and QCD.


\subsection{Scalar field theory}

Consider a scalar field theory where the Lagrangian depends only
on a scalar field $\phi$ and its derivative $\partial_\mu\phi$.
It is helpful to note the following variations:
\begin{subequations}
  \begin{align}
    \Delta \phi
    &=
    0
    \\
    \Delta (\partial_\mu \phi)
    &=
    -
    (\partial_\mu \dx^\nu)
    (\partial_\nu \phi)
    \,.
  \end{align}
\end{subequations}
Using the definition of the coefficients $\mathscr{D}^{\mu\nu}$
in Eq.~(\ref{eqn:D}):
\begin{subequations}
  \label{eqn:D:scalar}
  \begin{align}
    \mathscr{D}^{\mu\nu}[\phi]
    &=
    0
    \\
    \mathscr{D}^{\mu\nu}[\partial\phi]
    &=
    \frac{\partial\Lag}{\partial(\partial_\mu\phi)}
    (\partial^\nu \phi)
    \,.
  \end{align}
\end{subequations}
Using Eq.~(\ref{eqn:emt:general}) thus gives:
\begin{align}
  T^{\mu\nu}
  =
  \frac{\partial\Lag}{\partial(\partial_\mu\phi)}
  (\partial^\nu \phi)
  -
  g^{\mu\nu}
  \Lag
  \,,
\end{align}
which in this case coincides with the canonical EMT obtained
through Noether's first theorem.


\subsection{Covariant vector field theory}

Consider a vector field theory,
in which the Lagrangian depends only on a covariant
vector field $A_\mu$ and its derivative $\partial_\mu A_\rho$.
From Eq.~(\ref{eqn:variations}), note in particular:
\begin{subequations}
  \begin{align}
    \Delta A_\mu
    &=
    -
    (\partial_\mu \dx^\nu) A_\nu
    \\
    \Delta (\partial_\mu A_\rho)
    &=
    -
    (\partial_\mu \dx^\nu)
    (\partial_\nu A_\rho)
    -
    (\partial_\rho \dx^\nu)
    (\partial_\mu A_\nu)
    \,.
  \end{align}
\end{subequations}
Using the definition of the coefficients $\mathscr{D}^{\mu\nu}$
in Eq.~(\ref{eqn:D}):
\begin{subequations}
  \label{eqn:D:vector}
  \begin{align}
    \mathscr{D}^{\mu\nu}[A]
    &=
    \frac{\partial\Lag}{\partial A_\mu}
    A^\nu
    \\
    \mathscr{D}^{\mu\nu}[\partial A]
    &=
    \frac{\partial\Lag}{\partial(\partial_\mu A_\rho)}
    (\partial^\nu A_\rho)
    +
    \frac{\partial\Lag}{\partial(\partial_\rho A_\mu)}
    (\partial_\rho A^\nu)
    \,.
  \end{align}
\end{subequations}
Using Eq.~(\ref{eqn:emt:general}) gives:
\begin{align}
  \label{eqn:emt:vector}
  T^{\mu\nu}
  =
  \frac{\partial\Lag}{\partial(\partial_\mu A_\rho)}
  (\partial^\nu A_\rho)
  +
  \frac{\partial\Lag}{\partial(\partial_\rho A_\mu)}
  (\partial_\rho A^\nu)
  +
  \frac{\partial\Lag}{\partial A_\mu}
  A^\nu
  -
  g^{\mu\nu}
  \Lag
  \,,
\end{align}
which does not coincide with the canonical EMT found through
the first theorem.
In fact, only the first and last of these terms is present in the canonical EMT,
with the other two terms apparently being a correction to this.
The result obtained here is in agreement with the results of
Refs.~\cite{Felsager:1981iy,Munoz:1996wp,GamboaSaravi:2002vos,GamboaSaravi:2003aq,Montesinos:2006th}.

Let us consider the free electromagnetic field specifically.
The four-potential $A_\mu$ is a covariant rather than contravariant vector field,
since the covariant derivative $D_\mu \phi(x)$ should transform the same way
as $\partial_\mu \phi(x)$.
The Lagrangian is given by:
\begin{align}
  \Lag
  =
  - \frac{1}{4} F_{\mu\nu} F^{\mu\nu}
\end{align}
where $F_{\mu\nu} = \partial_\mu A_\nu - \partial_\nu A_\mu$.
Using the formula in Eq.~(\ref{eqn:emt:vector}),
the EMT is right away:
\begin{align}
  T^{\mu\nu}
  =
  F^{\mu\rho} F_{\rho}^{\phantom{\rho}\nu}
  +
  \frac{1}{4} g^{\mu\nu} F^2
  \,.
\end{align}
This result is identical to the Belinfante-improved EMT~\cite{Belinfante:1939emt},
and is the result widely accepted as the physical EMT of
electromagnetism~\cite{Jackson:1998nia}.
Noether's second theorem gave this result automatically,
without the need for an improvement procedure.


\subsection{Spinor field theory}

Let's consider a theory of a free fermion field next.
The Lagrangian depends on a single spinor field $\psi$,
along with its conjugate $\bar\psi$ and the derivatives of both fields.
As is standard, the conjugate field is treated as an independent field
with respect to variations.

The variations $\Delta \psi$ etc.\ are needed to proceed.
These can be obtained under the combined requirements that
$\bar\psi \psi$ transforms as a scalar field,
$\bar\psi \gamma^\mu \psi$ transforms as a contravariant vector field, and so on,
as well as that local translations of the form
$\dx^\mu(x) = \epsilon\omega^{\mu}_{\phantom{\mu}\nu}x^\nu$
reproduce the known behavior of spinors under infinitesimal
Lorentz transforms.
Transformations which satisfy these requirements are:
\begin{subequations}
  \label{eqn:variations:spinor}
  \begin{align}
    \Delta \psi
    &=
    \frac{i}{4}
    (\partial_\alpha \dx_\beta)
    \sigma^{\alpha\beta}
    \psi
    \\
    \Delta \bar\psi
    &=
    -
    \frac{i}{4}
    (\partial_\alpha \dx_\beta)
    \bar\psi
    \sigma^{\alpha\beta}
    \\
    \Delta (\partial_\mu\psi)
    &=
    \frac{i}{4}
    (\partial_\alpha \dx_\beta)
    \sigma^{\alpha\beta}
    (\partial_\mu \psi)
    -
    (\partial_\mu \dx^\lambda)
    (\partial_\lambda \psi)
    \\
    \Delta (\partial_\mu \bar\psi)
    &=
    -
    \frac{i}{4}
    (\partial_\alpha \dx_\beta)
    (\partial_\mu \bar\psi)
    \sigma^{\alpha\beta}
    -
    (\partial_\mu \dx^\lambda)
    (\partial_\lambda \bar\psi)
    \,,
  \end{align}
\end{subequations}
where $\sigma^{\alpha\beta} = \frac{i}{2}[\gamma^\alpha,\gamma^\beta]$.
It's worth remarking these are consistent with the pattern
found in Eq.~(\ref{eqn:delta:lie}),
where variations are related to Lie derivatives,
if compared to the spinor Lie derivative results
found by Kossmann~\cite{Kosmann:1971}.

Using the definition of the coefficients $\mathscr{D}^{\mu\nu}$
in Eq.~(\ref{eqn:D}):
\begin{subequations}
  \label{eqn:D:spinor}
  \begin{align}
    \mathscr{D}^{\mu\nu}[\psi]
    &=
    - \frac{i}{4}
    \frac{\partial\Lag}{\partial\psi}
    \sigma^{\mu\nu}
    \psi
    \\
    \mathscr{D}^{\mu\nu}[\bar\psi]
    &=
    \phantom{-} \frac{i}{4}
    \bar\psi
    \sigma^{\mu\nu}
    \frac{\partial\Lag}{\partial\bar\psi}
    \\
    \mathscr{D}^{\mu\nu}[\partial\psi]
    &=
    \frac{\partial\Lag}{\partial(\partial_\mu\psi)}
    (\partial^\nu \psi)
    -
    \frac{i}{4}
    \frac{\partial\Lag}{\partial(\partial_\rho\psi)}
    \sigma^{\mu\nu}
    (\partial_\rho \psi)
    \\
    \mathscr{D}^{\mu\nu}[\partial\bar\psi]
    &=
    (\partial^\nu \bar\psi)
    \frac{\partial\Lag}{\partial(\partial_\mu\bar\psi)}
    +
    \frac{i}{4}
    (\partial_\rho \bar\psi)
    \sigma^{\mu\nu}
    \frac{\partial\Lag}{\partial(\partial_\rho\bar\psi)}
    \,.
  \end{align}
\end{subequations}
Using Eq.~(\ref{eqn:emt:general}) with $\mathscr{D}^{\mu\nu}$
as the sum of these coefficients gives:
\begin{subequations}
  \label{eqn:emt:spinor}
  \begin{align}
    T^{\mu\nu}
    &=
    T^{\mu\nu}_{\mathrm{scl}}
    +
    T^{\mu\nu}_{\mathrm{spin}}
    \\
    T^{\mu\nu}_{\mathrm{scl}}
    &=
    \frac{\partial \Lag}{\partial(\partial_\mu\psi)}
    (\partial^\nu \psi)
    +
    (\partial^\nu \bar\psi)
    \frac{\partial \Lag}{\partial(\partial_\mu\bar\psi)}
    -
    g^{\mu\nu} \Lag
    \\
    T^{\mu\nu}_{\mathrm{spin}}
    &=
    -
    \frac{i}{4}
    \left\{
      \frac{\partial \Lag}{\partial \psi}
      \sigma^{\mu\nu}
      \psi
      -
      \bar\psi
      \sigma^{\mu\nu}
      \frac{\partial \Lag}{\partial \bar\psi}
      +
      \frac{\partial \Lag}{\partial(\partial_\rho\psi)}
      \sigma^{\mu\nu}
      (\partial_\rho\psi)
      -
      (\partial_\rho\bar\psi)
      \sigma^{\mu\nu}
      \frac{\partial \Lag}{\partial(\partial_\rho\bar\psi)}
      \right\}
    \,,
  \end{align}
\end{subequations}
where I have decomposed the EMT into a ``scalar'' (scl) piece
and a spin correction (spin) piece.
The ``scalar'' piece corresponds to what the result would be if the spinor field
transformed as a scalar.
The primary reason for performing the separation is as a calculation aid:
these pieces can be calculated separately and then combined.

For the Dirac Lagrangian in particular:
\begin{align}
  \Lag
  =
  \bar\psi
  \left( \frac{i}{2}\overleftrightarrow{\slashed{\partial}} - m \right)
  \psi
  \,,
\end{align}
where
$f \overleftrightarrow{\partial}_\mu g = f (\partial_\mu g) - (\partial_\mu f) g$.
The ``scalar'' and spin pieces of the
Dirac EMT are:
\begin{subequations}
  \begin{align}
    T^{\mu\nu}_{\mathrm{scl}}
    &=
    \frac{i}{2} \bar\psi \gamma^\mu \overleftrightarrow{\partial}^\nu \psi
    -
    g^{\mu\nu}
    \bar\psi
    \left( \frac{i}{2}\overleftrightarrow{\slashed{\partial}} - m \right)
    \psi
    \\
    T^{\mu\nu}_{\mathrm{spin}}
    &=
    \frac{1}{8}
    \Big\{
      \bar\psi
      [\gamma^\sigma, \sigma^{\mu\nu}]
      (\partial_\sigma\psi)
      -
      (\partial_\sigma\bar\psi)
      [\gamma^\sigma, \sigma^{\mu\nu}]
      \psi
      \Big\}
    \,.
  \end{align}
\end{subequations}
To proceed, the following identity is helpful:
\begin{align}
  \label{eqn:id:gs}
  [\gamma^\sigma, \sigma^{\mu\nu}]
  =
  2i( g^{\sigma\mu} \gamma^\nu - g^{\sigma\nu} \gamma^\mu )
  \,.
\end{align}
The spin correction piece can then be written:
\begin{align}
  T^{\mu\nu}_{\mathrm{spin}}
  &=
  \frac{i}{4}
  \bar\psi
  \Big(
  \gamma^\nu \overleftrightarrow{\partial}^\mu
  -
  \gamma^\mu \overleftrightarrow{\partial}^\nu
  \Big)
  \psi
  \,,
\end{align}
and adding both pieces together gives:
\begin{align}
  \label{eqn:emt:dirac}
  T^{\mu\nu}
  &=
  \frac{i}{4} \bar\psi \gamma^{\{\mu} \overleftrightarrow{\partial}^{\nu\}} \psi
  -
  g^{\mu\nu}
  \bar\psi
  \left( \frac{i}{2}\overleftrightarrow{\slashed{\partial}} - m \right)
  \psi
  \,,
\end{align}
where the brackets signify symmetrization over the indices,
e.g., $a^{\{\mu}b^{\nu\}} = a^\mu b^\nu + a^\nu b^\mu$.
Just as in the covariant vector case,
this is identical to the Belinfante EMT~\cite{Belinfante:1939emt},
but was obtained directly through Noether's second theorem.


\section{Application to physical gauge theories}
\label{sec:gauge}

I will now proceed to the ultimate purpose of this work:
obtaining the energy-momentum tensors for
quantum electrodynamics and quantum chromodynamics
using Noether's second theorem.
These theories are considered in a form that is used for actual field theoretic
calculations, which includes gauge-fixing terms needed to make the quantized
theory well-defined.


\subsection{Quantum electrodynamics with photon mass}

The QED Lagrangian is defined to contain a non-zero photon mass $\mu$
which regulates infrared divergences~\cite{Proca:1936fbw,Stueckelberg:1938hvi},
as well as a gauge-fixing term that makes the quantized theory
well-defined.
I use the Gupta-Bleuler
formalism~\cite{Gupta:1949rh,Bleuler:1950cy}
in particular for the gauge-fixing terms
(although the Nakanishi-Lautrup
formalism~\cite{Lautrup:1967zz,Nakanishi:1977gt,Nakanishi:1978zx,Nakanishi:1978np}
exists as an alternative).

The QED Lagrangian is given by:
\begin{align}
  \Lag_{\mathrm{QED}}
  =
  \bar\psi
  \left(
  \frac{i}{2}
  \overleftrightarrow{\slashed{D}}
  -
  m
  \right)
  \psi
  -
  \frac{1}{4} F_{\mu\nu} F^{\mu\nu}
  +
  \frac{1}{2} \mu^2 A_\mu A^\mu
  -
  \frac{\lambda}{2} (\partial_\mu A^\mu)^2
  \,.
\end{align}
where
$f \overleftrightarrow{D}_\mu g = f (D_\mu g) - (D_\mu f) g$,
and where the gauge-covariant derivative takes the forms:
\begin{subequations}
  \begin{align}
    \overrightarrow{D}_\mu \psi
    &=
    \overrightarrow{\partial}_\mu \psi
    +
    ie A_\mu \psi
    \\
    \bar\psi \overleftarrow{D}_\mu
    &=
    \bar\psi \overleftarrow{\partial}_\mu
    -
    ie \bar\psi A_\mu
    \,.
  \end{align}
\end{subequations}
The Lagrangian depends on the spinor fields $\psi$ and $\bar\psi$,
the covariant vector field $A_\mu$, and the derivatives of all three fields.
The results for the $\mathscr{D}^{\mu\nu}$ coefficients
defined in Eq.~(\ref{eqn:D}) are:
\begin{subequations}
  \begin{align}
    \mathscr{D}^{\mu\nu}[A]
    &=
    - e\bar\psi \gamma^\mu A^\nu \psi
    +
    \mu^2 A^\mu A^\nu
    \\
    \mathscr{D}^{\mu\nu}[\partial A]
    &=
    F^{\mu\rho}
    F_{\rho}^{\phantom{\rho}\nu}
    -
    \lambda
    (\partial^{\{\mu} A^{\nu\}})
    (\partial_\rho A^\rho)
    \\
    \mathscr{D}^{\mu\nu}[\psi]
    &=
    -
    \frac{1}{8} (\partial_\rho \bar\psi) \gamma^\rho \sigma^{\mu\nu} \psi
    +
    \frac{i}{4} e \bar\psi \slashed{A} \sigma^{\mu\nu} \psi
    +
    \frac{i}{4} m \bar\psi \sigma^{\mu\nu} \psi
    \\
    \mathscr{D}^{\mu\nu}[\partial \psi]
    &=
    \frac{i}{2}
    \bar\psi \gamma^\mu
    (\partial^\nu \psi)
    +
    \frac{1}{8}
    \bar\psi \gamma^\rho
    \sigma^{\mu\nu} (\partial_\rho \psi)
    \\
    \mathscr{D}^{\mu\nu}[\bar\psi]
    &=
    -
    \frac{1}{8} \bar\psi \sigma^{\mu\nu}
    \gamma^\rho (\partial_\rho \psi)
    -
    \frac{i}{4}
    e
    \bar\psi \sigma^{\mu\nu} \slashed{A} \psi
    -
    \frac{i}{4}
    m
    \bar\psi \sigma^{\mu\nu} \psi
    \\
    \mathscr{D}^{\mu\nu}[\partial \bar\psi]
    &=
    - \frac{i}{2} (\partial^\nu\bar\psi) \gamma^\mu \psi
    + \frac{1}{8} (\partial_\rho \bar\psi) \sigma^{\mu\nu}
    \gamma^\rho \psi
    \,.
  \end{align}
\end{subequations}
Adding the spinor pieces together gives:
\begin{align}
  \mathscr{D}^{\mu\nu}[ \psi ]
  +
  \mathscr{D}^{\mu\nu}[ \partial\psi ]
  +
  \mathscr{D}^{\mu\nu}[ \bar\psi ]
  +
  \mathscr{D}^{\mu\nu}[ \partial\bar\psi ]
  &=
  \frac{i}{2} \bar\psi
  \gamma^\mu \overleftrightarrow{\partial}^\nu
  \psi
  +
  \frac{i}{4}
  \bar\psi
  \Big(
  \gamma^\nu \overleftrightarrow{D}^\mu
  -
  \gamma^\mu \overleftrightarrow{D}^\nu
  \Big)
  \psi
  \,.
\end{align}
Adding these to the photon pieces and using
Eq.~(\ref{eqn:emt:general}) for the EMT gives:
\begin{align}
  \label{eqn:emt:qed}
  T^{\mu\nu}_{\mathrm{QED}}
  =
  \frac{i}{4} \bar\psi
  \gamma^{\{\mu} \overleftrightarrow{D}^{\nu\}}
  \psi
  + F^{\mu\sigma} F_{\sigma}^{\phantom{\sigma}\nu}
  + \mu^2 A^\mu A^\nu
  - \lambda (\partial\cdot A) \partial^{\{\mu} A^{\nu\}}
  -
  g^{\mu\nu} \Lag_{\mathrm{QED}}
  \,.
\end{align}
This EMT is symmetric in its indices,
and in the case $\mu=0$ and $\lambda=0$
is also clearly gauge invariant.
The result is moreover consistent with the previous
Proca theory result in Ref.~\cite{Montesinos:2006th} (taking $\lambda=0$)
and with the result (found via the Belinfante improvement procedure)
for massive QED with gauge-fixing found in Ref.~\cite{Embacher:1986pt}.


\subsection{Quantum chromodynamics}

Lastly, I will use Noether's second theorem
to obtain the EMT for quantum chromodynamics.
For the quantized theory to be well-defined,
there must not only be a gauge-fixing term,
but also gauge-compensating terms that subtract off
contributions from counting unphysical gluon modes.
This is done by introducing Faddeev-Popov ghosts~\cite{Faddeev:1967fc},
which are Lorentz scalar fields that are quantized with fermi statistics.
The full QCD Lagrangian can be
written~\cite{Becchi:1974md,Becchi:1975nq,Tyutin:1975qk,Kugo:1979gm}\footnote{
  Multiple formulations of the QCD Lagrangian exist in the literature,
  which are equivalent when equations of motion are applied.
  I am using the Lagrangian of Kugo and Ojima~\cite{Kugo:1979gm}
  because it is invariant under BRST transformations,
  while other formulations are only invariant up to a total derivative.
}:
\begin{align}
  \label{eqn:lagrangian:QCD}
  \Lag_{\mathrm{QCD}}
  =
  \sum_q
  \bar{q}
  \left(
  \frac{i}{2}
  \overleftrightarrow{\slashed{\partial}}
  +
  g\slashed{A}_a T^a
  -
  m_q
  \right)
  q
  -
  \frac{1}{4} F^a_{\mu\nu} F_a^{\mu\nu}
  - (\partial_\mu B_a) A^\mu_a
  +
  \frac{\alpha_0}{2} B_a^2
  -
  i (\partial_\mu \bar{c}^a) (D^\mu_{ab} c^b)
  \,,
\end{align}
where $B_a$ are Lagrange multiplier fields and $c_a$ and $\bar{c}_a$
are the Faddeev-Popov ghosts.
Unlike with QED, a gluon mass cannot be introduced to QCD
without breaking renormalizability~\cite{Salam:1962aq,Boulware:1970zc},
so I will not consider a gluon mass term here.
The different representations of the gauge-covariant derivative are:
\begin{subequations}
  \begin{align}
    \overrightarrow{D}_\mu q
    &=
    \overrightarrow{\partial_\mu} q
    -
    ig A^a_\mu T_a q
    \\
    \bar{q} \overleftarrow{D}_\mu
    &=
    \bar{q} \overleftarrow{\partial_\mu}
    +
    ig \bar{q} A^a_\mu T_a
    \\
    D_\mu^{ab} c^b
    &=
    \Big(
    \delta_{ab} \partial_\mu
    + g f_{acb} A_\mu^c
    \Big)
    c^b
    \,,
  \end{align}
\end{subequations}
and the gluon field strength tensor is:
\begin{align}
  F_{\mu\nu}^a
  =
  \partial_\mu
  A_\nu^a
  -
  \partial_\nu
  A_\mu^a
  +
  g f_{abc} A_\mu^b A_\nu^c
  \,.
\end{align}
Here, $T_a$ are the generators of the color $\mathfrak{su}(3,\mathbb{C})$
algebra and $f_{abc}$ are the totally antisymmetric structure constants
defined by:
\begin{align}
  [T_a, T_b] = i f_{abc} T_c
  \,.
\end{align}
Because gauge-fixing terms are present,
the QCD Lagrangian is not gauge-invariant.
However, it is invariant instead under the larger
Becchi-Rouet-Stora-Tyutin (BRST) transformation
group~\cite{Becchi:1974md,Becchi:1975nq,Tyutin:1975qk},
which allows for simple proofs of Ward identities
and ensures renormalizability of the theory~\cite{Collins:1984xc}.
The infinitesimal actions of these transformations on the fields
in the QCD Lagrangian are~\cite{Becchi:1974md,Becchi:1975nq,Tyutin:1975qk,Kugo:1979gm}:
\begin{subequations}
  \begin{align}
    \delta A_\mu^a
    &=
    \delta\lambda
    D_\mu^{ab} c^b
    \\
    \delta c^a
    &=
    -
    \frac{1}{2}
    \delta\lambda
    g f_{abc} c_b c_c
    \\
    \delta c^a
    &=
    i
    \delta\lambda
    B^a
    \\
    \delta B^a
    &=
    0
    \\
    \delta q
    &=
    i T_a \delta\lambda c_a q
    \,,
  \end{align}
\end{subequations}
where $\delta\lambda$ is a Grassmann number valued parameter.
Since $c_a$ is also Grassmann number valued, this effectively
transforms the quark and gluon fields in the same manner as
an infinitesimal gauge transformation.
It is also straightforward to see that such a transformation
leaves Eq.~(\ref{eqn:lagrangian:QCD}) invariant.
(For a full explication of the formalism,
see Ref.~\cite{Kugo:1979gm}.)

It is natural, therefore, to expect the EMT of QCD to be BRST-invariant.
Additionally, what remains of the EMT when gauge-fixing and ghost terms
are dropped should be gauge-invariant too.

Using Eq.~(\ref{eqn:D}),
the non-trivial $\mathscr{D}^{\mu\nu}$ coefficients evaluate to:
\begin{subequations}
  \begin{align}
    \mathscr{D}^{\mu\nu}[{q}]
    &=
    -
    \frac{1}{8} (\partial_\rho \bar{q}) \gamma^\rho \sigma^{\mu\nu} {q}
    -
    \frac{i}{4} g \bar{q} \slashed{A}_a T^a \sigma^{\mu\nu} {q}
    +
    \frac{i}{4} m \bar{q} \sigma^{\mu\nu} {q}
    \\
    \mathscr{D}^{\mu\nu}[\partial {q}]
    &=
    \frac{i}{2}
    \bar{q} \gamma^\mu
    (\partial^\nu {q})
    +
    \frac{1}{8}
    \bar{q} \gamma^\rho
    \sigma^{\mu\nu} (\partial_\rho {q})
    \\
    \mathscr{D}^{\mu\nu}[\bar{q}]
    &=
    -
    \frac{1}{8} \bar{q} \sigma^{\mu\nu}
    \gamma^\rho (\partial_\rho {q})
    +
    \frac{i}{4}
    g
    \bar{q} \sigma^{\mu\nu} \slashed{A}^a T_a {q}
    -
    \frac{i}{4}
    m
    \bar{q} \sigma^{\mu\nu} {q}
    \\
    \mathscr{D}^{\mu\nu}[\partial \bar{q}]
    &=
    - \frac{i}{2} (\partial^\nu\bar{q}) \gamma^\mu {q}
    + \frac{1}{8} (\partial_\rho \bar{q}) \sigma^{\mu\nu}
    \gamma^\rho {q}
    \\
    \mathscr{D}^{\mu\nu}[A]
    &=
    \sum_q
    g \bar q \gamma^\mu A^\nu T_a q
    + g f_{abc} F_b^{\mu\rho} A^c_\rho A^\nu_a
    - ig f_{abc} (\partial_\mu \bar{c}^c) A^\nu_a c^b
    -
    (\partial^\mu B_a) A^\nu_a
    \\
    \mathscr{D}^{\mu\nu}[\partial A]
    &=
    F^{\mu\rho}_a
    ( \partial_\rho A^\nu_a - \partial^\nu A_\rho^a )
    \\
    \mathscr{D}^{\mu\nu}[\partial B]
    &=
    -
    A^\mu_a (\partial^\nu B_a)
    \\
    \mathscr{D}^{\mu\nu}[\partial c]
    &=
    -i (\partial^\nu \bar{c}^a) (\partial^\mu c^a)
    \\
    \mathscr{D}^{\mu\nu}[\partial \bar{c}]
    &=
    - i (\partial^\nu \bar{c}^a)(D^\mu_{ab} c^b)
    \,.
  \end{align}
\end{subequations}
The remaining coefficients are zero.
Summing the quark contributions and using Eq.~(\ref{eqn:id:gs}) gives:
\begin{subequations}
  \begin{align}
    \mathscr{D}^{\mu\nu}[ {q} ]
    +
    \mathscr{D}^{\mu\nu}[ \partial{q} ]
    +
    \mathscr{D}^{\mu\nu}[ \bar{q} ]
    +
    \mathscr{D}^{\mu\nu}[ \partial\bar{q} ]
    &=
    \frac{i}{2} \bar q
    \gamma^\mu \overleftrightarrow{\partial}^\nu
    q
    +
    \frac{i}{4}
    \bar q
    \Big(
    \gamma^\nu \overleftrightarrow{D}^\mu
    -
    \gamma^\mu \overleftrightarrow{D}^\nu
    \Big)
    q
    \,,
  \end{align}
\end{subequations}
Adding all of the contributions together gives an EMT
that is symmetric in its indices and BRST invariant:
\begin{align}
  \label{eqn:emt:qcd}
  T_{\mathrm{QCD}}^{\mu\nu}
  =
  \sum_q \frac{i}{4} \bar q \gamma^{\{\mu} \overleftrightarrow{D}^{\nu\}} q
  + F_a^{\mu\rho} F_{\rho}^{a\,\nu}
  - A_a^{\{\mu} \partial^{\nu\}} B_a
  -i (D^{\{\mu} c) (\partial^{\nu\}} \bar{c})
  -
  g^{\mu\nu} \Lag_{\mathrm{QCD}}
  \,.
\end{align}
Demonstrating BRST invariance is algebraically involved,
but is no more difficult---and in fact, is essentially identical---to
proving BRST invariance of
the Lagrangian in Eq.~(\ref{eqn:lagrangian:QCD}) itself.
If the physical sector alone is considered
(i.e., the ghost and gauge-fixing fields dropped),
then the remaining terms are locally gauge invariant.
The necessary symmetries are therefore automatically satisfied from applying
Noether's second theorem alone, without the need to apply an
ad hoc improvement procedure.

It should be noted that Eq.~(\ref{eqn:emt:qcd}) agrees with the
EMT found using the Belinfante improvement procedure in
Refs.~\cite{Kugo:1979gm,Leader:2013jra}.


\section{Discussion and Conclusions}
\label{sec:conclusion}

Ultimately, Noether's first and second theorems both imply conserved currents
that can be identified as an energy-momentum tensor.
The first theorem identifies the EMT as the conserved current associated
with global spacetime translations,
while the second theorem identifies it as a conserved current
associated with \emph{local} spacetime translations.

In this work, I derived the EMT following from Noether's second theorem
and local translation invariance, for both
quantum electrodynamics in Eq.~(\ref{eqn:emt:qed})
and quantum chromodynamics in Eq.~(\ref{eqn:emt:qcd}).
The EMT operator for both theories is symmetric in its indices
and invariant under the expected symmetries (gauge or BRST invariance).
It is identical for the Belinfante-improved EMT in both cases,
as well as to the EMT of general relativity that one would obtain through metric
variations~\cite{Belitsky:2005qn}.

Although the form of the EMT thus derived is not new,
the ability to derive it directly from one of Noether's theorems---without
need for an improvement procedure---puts this form of the EMT
on surer theoretical ground.
Additionally, the decomposition of the EMT into the
$\mathscr{D}^{\mu\nu}$ coefficients defined in Eq.~(\ref{eqn:D})
may provide a new perspective on how to decompose components of the EMT
into quark and gluon contributions, at least for components for which
$-g^{\mu\nu}\Lag$ is zero.
The $\mathscr{D}^{\mu\nu}$ terms for each field arise directly from its
variation under local translations,
and should accordingly relate to generators of translations and rotations
for the field in question.
However, since the $-g^{\mu\nu}\Lag$ term in the EMT arises from the Jacobian
of a coordinate transformation, the method explored here does not provide
a clear way to decompose it into quark and gluon contributions.

A possible avenue of future research is to study the momentum and spin
decompositions suggested by this method of derivation,
and to compare these with the well-known Jaffe-Manohar~\cite{Jaffe:1989jz}
and Ji~\cite{Ji:1996ek} decompositions.

\begin{acknowledgments}
  I would like to thank
  Ian Clo\"et and Gerald Miller
  for helpful discussions that contributed to this work,
  and would additionally like to thank C\'edric Lorc\'e,
  Boris Kosyakov, and Peter Rau
  for insightful comments and technical scrutiny that
  improved the manuscript in revision.
  This work was supported by the U.S.\ Department of Energy
  Office of Science, Office of Nuclear Physics under Award Number
  DE-FG02-97ER-41014.
\end{acknowledgments}


\bibliography{references.bib}

\end{document}